\documentclass[conference]{IEEEtran}
\usepackage{xcolor}
\usepackage{amssymb}
\usepackage{epsfig, subfigure, amsmath, amssymb}
\usepackage{wrapfig,lipsum,booktabs}

\usepackage{amsthm, algorithm, algpseudocode}
\usepackage{epstopdf}
\usepackage{balance}
\usepackage{color}
\usepackage{url}
\usepackage{pifont}
\usepackage{paralist}
\usepackage{commath}
\usepackage{enumitem}
\usepackage{balance}
\usepackage{comment}
\usepackage{multirow}
\usepackage{upgreek}

\usepackage{txfonts}
\usepackage[T1]{fontenc}

\definecolor{red}{rgb}{1,0,0}

\definecolor{blue}{rgb}{0,0,1}

\newcommand{\ie}{{ i. e.}, }
\newcommand{\etal}{{ et al.}, }

\AtBeginDocument{%
  \providecommand\BibTeX{{%
    \normalfont B\kern-0.5em{\scshape i\kern-0.25em b}\kern-0.8em\TeX}}}

\begin{document}


\title{False Relay Operation Attacks in \\ Power Systems with High Renewables}

 \author{%
\IEEEauthorblockN{Mohamadsaleh Jafari, Md Hassan Shahriar, Mohammad Ashiqur Rahman, and Sumit Paudyal}
\IEEEauthorblockA{
Department of Electrical and Computer Engineering, Florida International University, USA\\
Emails: mjafari@fiu.edu, mshah068@fiu.edu, marahman@fiu.edu,   spaudyal@fiu.edu} 
}
\maketitle

\begin{abstract}
Load-generation balance and system inertia are essential for maintaining frequency in power systems. Power grids are equipped with Rate-of-Change-of-Frequency ($ROCOF$) and Load Shedding ($LS$) relays in order to keep load-generation balance. With the increasing penetration of renewables, the inertia of the power grids is declining, which results in a faster drop in system frequency in case of load-generation imbalance. In this context, we analyze the feasibility of launching False Data Injection (FDI) in order to create  False Relay Operations (FRO), which we refer to as FRO attack, in the power systems with high renewables. 
We model the frequency dynamics of the power systems and corresponding FDI attacks, including the impact of parameters, such as synchronous generators'  inertia, and governors' time constant and droop, on the success of FRO attacks. We formalize the FRO attack as a Constraint Satisfaction Problem (CSP) and solve using Satisfiability Modulo Theories (SMT).
Our case studies show that power grids with renewables are more susceptible to FRO attacks and the inertia of synchronous generators plays a critical role in reducing the success of FRO attacks in the power grids. 
\end{abstract}


\maketitle

\section {Nomenclature}

\begin{tabbing}
    
    \=$\Delta f$\ \hspace{15pt} \= Change in frequency.\\
	\(\Delta P\)  \>\>  Total power imbalance. \\
	\(\Delta P^{gov}\)  \>\>  Change in  power due to governor's action. \\
	\(\Delta P^{a}\)  \>\>  Change in generator's setpoint due to  FDI attack. \\
	\(\Delta P^{sh}\)\>\> Shed load due to Load Shedding ($LS$) relay.\\
	\(\Delta P^{tg}\)\>\> Change of power  generation due to $ROCOF$.\\
	\(\Delta t\)\>\> Simulation time step. \\
	\(f\)    \>\>  Frequency. \\
	\(\overline{f}\)\>\> Frequency threshold for load shedding.\\ 
	\(\dot f\)\>\> Rate-of-Change-of-Frequency ($ROCOF$).\\ 
	\(\overline{\dot f}\)\>\> Threshold value of $ROCOF$. \\
	\(H\) \>\> Inertia constant of multi-machine System.\\ 
	\(M\)\>\> Number of cycles in $\dot f$ calculation.\\ 
	\(n\)  \>\>  Discrete time step. \\
	\(P^e\)  \>\>  Electrical power output of  generators. \\
	\(P^m\)  \>\>  Mechanical power input to generators. \\
	\(P^{sh}\)\>\> Load  shed by $LS$ relays at each time step.\\
	\(P^{tg}\)\>\> Power of tripped  generators by $ROCOF$ relays.\\
	\(R\)\>\> Droop of  the governors.\\ 
	\(T\)\>\> Time constant of  governors. \\
	\(t\)    \>\>  Time. \\
	
\end{tabbing}

\section{Introduction}
\label{Sec:Introduction}

A large part of power generation in bulk power systems is supplied by synchronous generators. However, these days, Distributed Energy Resources (DERs) are becoming an integral part of the power systems \cite{meProtection}. As the power grid evolves with increasing penetrations of inverter-based DERs such as solar photovoltaics (PV) and wind turbines, the inertia of the grid tends to decline \cite{tamrakar2017virtual, NREL}. 

Wind turbines and solar panels are equipped with several sensor measurements such as wind velocity, wind direction, and solar irradiance.   
These measurements are connected to the control center using wireless/wired technologies to facilitate analyzing and adjusting the power output of generators to maintain the grid frequency.  
Protective devices, such as Rate-of-Change-of-Frequency ($ROCOF$) and Load Shedding ($LS$) relays are equipped in the power system to maintain load-generation balance when frequency changes \cite{rocof2020,Tofis2017}. In the case of load-generation imbalance, the system frequency deviates from the nominal value, and if the frequency/$ROCOF$ goes beyond an acceptable range, these relays trip their corresponding generators/loads to keep the frequency within the range. 
An attacker can exploit this relay functionality by performing False Data Injection (FDI) attacks. 
The attacker may inject necessary false data into various DER sensor measurements causing a False Relay Operation (FRO) (i.e., false operation of an $LS$/$ROCOF$ relay) in the grid. For example, the attacker can inject false data into the air density and wind velocity measurements of a wind turbine to mislead the control center to perceive a power abundance in the network. Then, the operator of the control center adjusts the power output of synchronous generators. However, as the actual generation is less than the load, the frequency drops. If this attack continues, some generators may trip by $ROCOF$ relay operation or some loads may get disconnected by $LS$ relays. We name such exploitations as FRO attacks.

Various FDI attacks in power grids have been widely studied in the literature. The authors in \cite{rahman2019novel,kang2018false} presented an FDI attack on contingency analysis of the power system, where the target was created to mislead the operational cost only.
Chlela et al.~\cite{7741747} addressed the impacts of FDI cyber-attacks on critical microgrid control functions as well as the loss of load resulting from under-frequency load shedding. 
Zhang et al.~\cite{7249479} discussed security issues of a  dynamic microgrid partition process and investigated three different scenarios of FDI attacks against it. 
In \cite{6503599}, the authors analyzed FDI attacks with incomplete information, where the attacker has limited information about the network topology. 
The possible FDI vulnerabilities on an integrated Volt-VAr control is
studied in~\cite{6859265}. 
Teixeira et al.~\cite{7301476} studied the impact of FDI attacks on the measurement data and reference signals received by the voltage droop controllers in microgrids.
Liu\etal  \cite{9120330} focus on the continuous injection of time-varying false data and load redistribution in the system.
A pre-overload vulnerability graph approach is proposed in~\cite{9138481} to systematically assess, evaluate, and quantify the system vulnerability under a load redistribution FDI attack.
Some researchers studied the defense against the above-mentioned attacks. For example, a reachability analysis-based mechanism is presented in~\cite{beg2017detection} to detect FDI attacks. Saad et al.~\cite{9112234} presented an IoT-based cyber-physical system at mitigating FDI attacks. 

However, to the best of our knowledge, the current literature does not address the possibility of launching FDI attacks leading to FRO. In this context, this paper aims at studying the feasibility of FRO attacks in power systems. 
We represent the overall problem in a generic, formal manner by recognizing it as a Constraint Satisfaction Problem (CSP). We apply Satisfiability Modulo Theories (SMT), a powerful constraint satisfaction tool~\cite{WT-SMT}, to solve this CSP. 

The rest of the paper is organized as follows. In section \ref{Sec:FormalModeling}, we model the frequency behavior of the power system considering the change of the generator setpoints due to FDI, the generator's governor reaction, $LS$, and $ROCOF$ relay operations. In section \ref{Sec:CaseStudies}, we show case studies by considering different combinations of power system parameters and analyze their impacts on the success of FDI in launching FRO attacks. We conclude the paper in section \ref{Sec:Conclusion}. 

\vspace{-1pt}
\section{Formal Models}
\label{Sec:FormalModeling}

In this section, we present the formal models of the power system frequency dynamics and the synthesis of potential FRO attack vectors. An attack vector identifies a set of sensor measurements and necessary FDIs that can lead to a FRO attack. We also model the power grid relay operations and how the attack can percolate into this system leading to false activation of the relays. We specifically focus on $ROCOF$ and $LS$ relays.

\subsection{Power System Frequency Dynamics}


The system frequency response in a power grid can be determined using swing equation, and load/generation changes.  For multi-machine power systems, the swing equations can be equivalently represented in terms of center of inertia (COI) as, 
\begin{equation}
\frac{d\,f(t)}{dt}= \frac{1}{2H}\,{(P^m - P^e).}   
\label{e1}
\end{equation}

The above form of swing equation can be 
linearized for a multi-machine power system as follows  \cite{amraee2018probabilistic},
\begin{equation}
    \frac{d\,\Delta f(t)}{dt}=\frac{1}{2H}\,{\Delta P(t).}\, 
    \label{e5}
\end{equation}

\subsection{FDI Attacks on Grid Frequency Control}
\label{SubSec:Frequency_Control}

In this study, the FDI attack on measurements from DERs (e.g., wind velocity, solar irradiance)  will be perceived as generation change at the control center level. We assume that the attacker cannot directly change the generators' setpoints. Therefore, the attacker tries to launch FDI attacks on the DER measurements. These compromised measurements are sent to the control center to mislead the operator of the abundance or shortage of power in the network. Then, the operator sends new setpoints to the synchronous generators to address the issue, while unknowingly participating in the attacker's goal. We assume that the frequency regulation takes place mainly due to the primary frequency response and secondary frequency control actions.

We model the power imbalance as \cite{amraee2018probabilistic}, 
 %
\begin{equation}
    \begin{split}
     \Delta P_n=\Delta P_n^{gov}-\Delta P_n^{a}+\Delta P_n^{sh} - \Delta P_n^{g}.
     \label{e6}
\end{split}
\end{equation}

\begin{figure}[b]
\vspace{-14pt}
    \centering
    \includegraphics[scale=0.28, keepaspectratio=true]{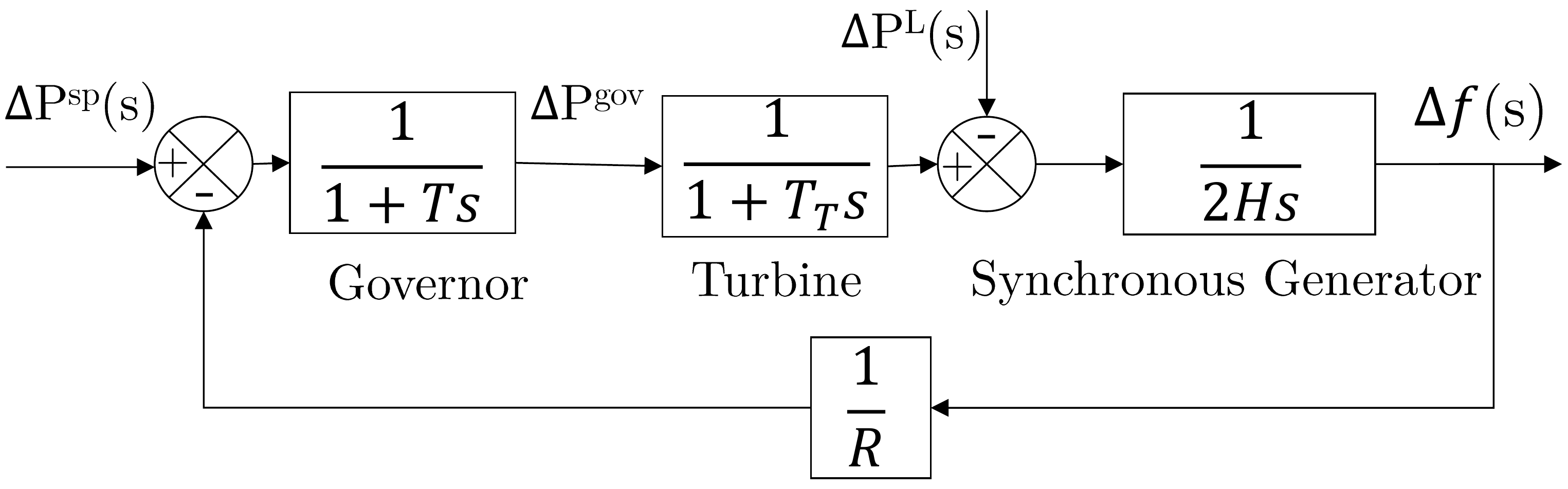}
    \caption{Model of generators' Equivalent Governor. }
    \label{fig:govenor}
\end{figure}

Governor operation ($\Delta P_n^{gov} \neq 0$), as shown in Fig.~\ref{fig:govenor},
changes the generators input power during any load-generation imbalances. 
The discretized dynamic response of the governor is modeled using the following \cite{amraee2018probabilistic}, 
%
\begin{equation}
     \Delta P_{n+1}^{gov}=
    \Delta P_{n}^{gov}+\frac{\Delta t}{T} \big(-\frac{\Delta f_n}{R}-\Delta P_{n}^{gov} \big), 
    \label{e9}
\end{equation}

We model the frequency resulting from the actions of governors, $ROCOF$, and $LS$ relays using the following,
\begin{align}
\label{del_f}
   &\Delta f_{n+1} = \frac{\Delta t}{4H} 
     \Big [ \Delta P ^{gov}_n ( 2-\frac{\Delta t}{T})- 2 \Delta P_n^{a} \nonumber  \\ 
&-\Delta f_n (\frac {\Delta t}{RT}- \frac{4H}{\Delta t}) - \Delta P^{tg}_{n+1} - \Delta P^{sh}_{n+1} \Big ],  \\  
& f_n = 1 + \Delta f_n,
\label{frequency}
\end{align}

If $f_n$ is less than $\overline{f}$, 
the total amount of shed load at time  $n+1$ is computed as,
\begin{equation}
\label{load_sheddig}
 \Delta P^{sh}_{n+1} = \Delta P^{sh}_n + P^{sh}, \quad \quad \forall f_n \leq \overline{f}
\end{equation}
To model the activation of $ROCOF$ relays, we compute $\dot f$ as follows \cite{alam2019},
\begin{equation}
\label{ROCOF}
    \dot f = \frac{1}{M} \sum_{n-M+1}^{n} \frac{\Delta f_n}{\Delta t}, 
\end{equation}
If $\dot f$  is greater than $\overline{\dot f}$,
the $ROCOF$ relay operates and disconnects the generator from the grid. This generator disconnection is modeled by, 
\begin{equation}
\label{del_P_SG_n+1}
 \Delta P^{tg}_{n+1} = \Delta P^{tg}_{n} + P^{tg},\quad \quad \forall  \dot f \geq \overline{\dot f}
\end{equation}

\vspace{2pt}
\section{Case Studies}
\label{Sec:CaseStudies}

\begin{figure}[b!]
    \begin{center}
\includegraphics[width=0.43\textwidth]{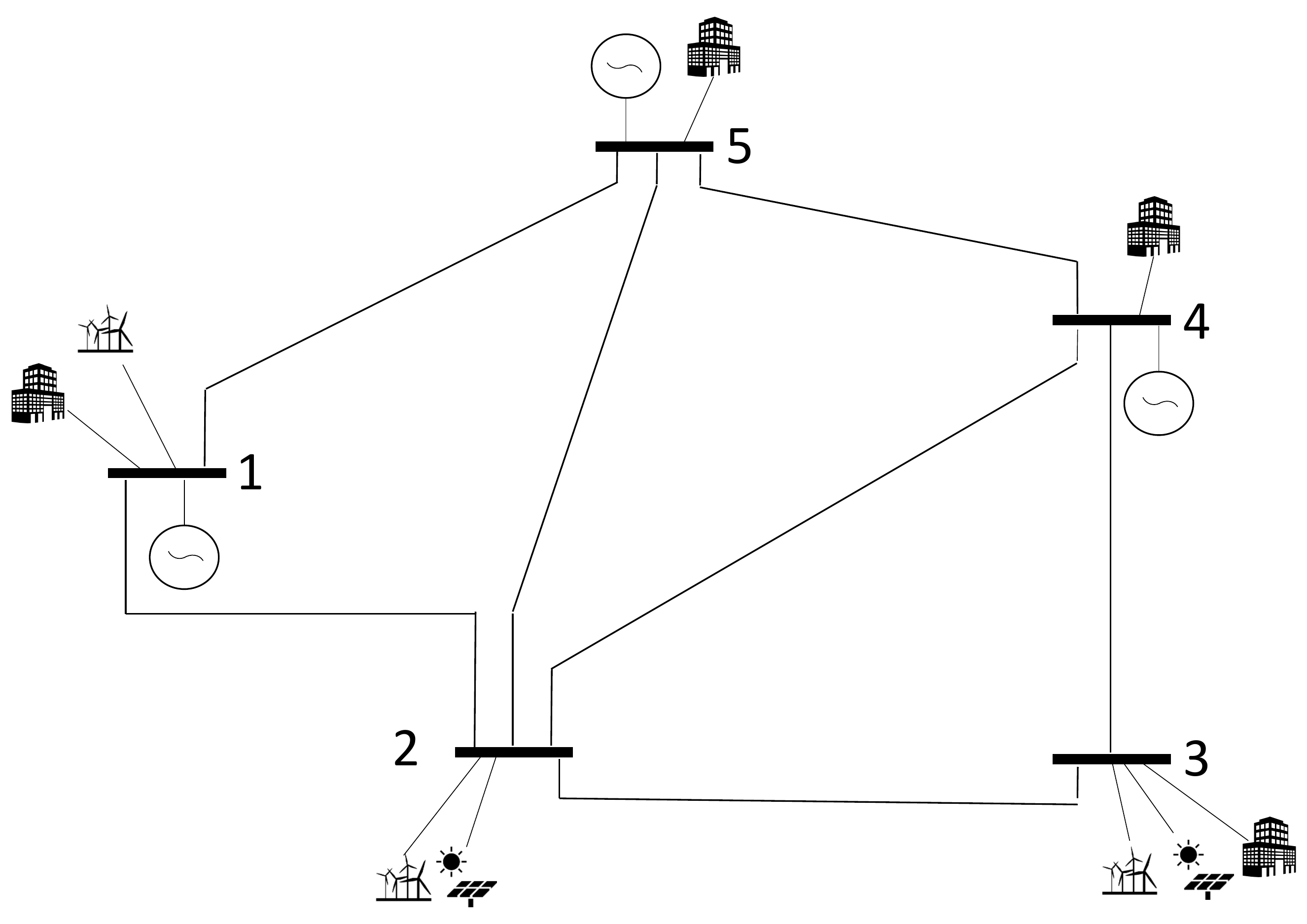}
   \vspace{-9pt}
    \caption{A 5-Bus Test Network for case studies.}
    \label{Fig_Case_Study_5Bus}
    \end{center} 
     \vspace{-9pt}
\end{figure}

In order to study the proposed method, we consider a 5-bus power system as shown in Fig.~\ref{Fig_Case_Study_5Bus} with three generators, three wind farms, two solar parks, and four bulk loads. The total amount of the load is 4.5 p.u., 3.0 p.u. of which is supplied by the generators (1.0 p.u. each), and 1.50 p.u. is supplied by the wind and solar generations. We consider a nominal frequency of 60~Hz.
All the loads are equipped with $LS$ relays with $\overline{f}$ = 59.5 Hz.
Generators are equipped with $ROCOF$ relays with $\overline{\dot f}$= 0.5,  0.6, and 1.2 Hz/s at buses 4, 5, and 1, respectively. The reason for considering different $\overline{\dot f}$ values for the relays is that not all of the generators in case of any $ROCOF$-type attack get disconnected simultaneously from the grid \cite{1645146}. The acceptable maximum $\overline{\dot f}$ is usually ranged between 0.5 and 1.2 Hz/s \cite{8332112}. The value of $M$
typically varies between 2 and 40 cycles \cite{Ten2008}. In our study,
we set $M$ to 6 cycles. Also, we assume that the FDI attack takes place only at a single time step (not a recurring attack attempt).
We encode the formalization presented in the previous section and solve it using Z3, which is an efficient SMT solver~\cite{WT-SMT}. During the execution of the FDI attack, the solution to the model returns either a successful or unsuccessful status.  
If the result is successful, it denotes that the SMT solver identified an attack vector that satisfies all the given constraints. On the other hand, unsuccessful status implies that there is no solution to the given problem (with the given attack constraints). In this work, an attack vector represents an assignment variable value for which the framework identifies a satisfiable solution. We run our experiments on an Intel Core i7 processor with 16 GB memory.
In order to evaluate the impact of power system parameters on the feasibility of FDI on FRO, we consider different combinations of $H$, $R$, and $T$ for the example power grid and run the proposed framework. We also consider the Threshold of Injection ($ToI$) and  Attackable DERs ($AD$) parameters in our studies. $ToI$ is the maximum percentage of the change that the attacker can make in measurements as false data and $AD$ is the percentage of attackable DERs (including wind turbines and solar panels). By attackable, we mean that the measurements in DERs are not secured and the attacker can access them.
\begin{table}[t]
\vspace{-15pt}
\caption{Power system parameters used for the case studies.}
    \label{tab:Evaluation parameters}
    \centering
        \begin{tabular}{|c|c|c|c|}
    \hline
     \textbf{Parameter} &  \multicolumn{3}{|c|}{\textbf{Case Study}}\\ \hline 
           & $C1$ & $C2$ & $C3$\\ \hline
    $R(pu)$  & 0.2 &  0.2  &  0.2, 0.4, 0.6, 0.8, 1.0 \\ \hline
    $T(s)$   & 0.2 &  0.2  &  0.2, 0.4, 0.6, 0.8, 1.0 \\ \hline 
    $AD\%$   & 20.0  &  20.0   &  20.0,  40.0,  60.0,  80.0,  100.0 \\ \hline
    $H(s)$   & 2.0 &  6.0  &  2.0, 4.0, 6.0, 8.0, 10  \\ \hline
    $ToI\%$  & 2.0 &  6.0  &  2.0, 4.0, 6.0, 8.0, 10  \\ \hline
    \end{tabular}
\end{table}
Table~\ref{tab:Evaluation parameters} 
shows different values of the example power grid parameters used to evaluate the performance of the proposed framework in different case studies. To generate a generic relationship among the parameters, we created 10,000 different combinations of these values and observed the successful attacks at launching FRO. 
Among these combinations, we pick two to show the frequency and $ROCOF$ behaviors of the power system. Then, we show a bigger picture of these parameters impact on attack success.

\begin{itemize}[leftmargin=*]

\item {Case 1 ($C1$):} 
Injecting false data into DERs' measurements, the attacker launches an attack of $\Delta P_1^a $ = 0.322 p.u. in the system. This attack convinces the operator to reduce the generators' setpoints to address this power abundance. Therefore, $\dot f$ sharply changes and reaches 0.510 Hz/s in 12 cycles which is greater than the $\overline{\dot f}$ of the generator at bus 4. Therefore, $ROCOF$ relay at bus 4 operates falsely and disconnects its corresponding generator from the grid ($P^{tg}$ = 1.0 p.u.). The frequency and $\dot f$ changes are shown in Table~\ref{Tab:Output} for $n$ = 0,...,12.

\begin{table}[b]
\vspace{-9pt}
\caption{Frequency and ROCOF behavior of the example power system for case studies $C1$ and $C2$.}
    \label{Tab:Output}
    \centering
    \begin{tabular}{|c|c|c|c|c|} \hline
    $n$ &\multicolumn{2}{|c|}{$\dot f (Hz/s)$} & \multicolumn{2}{|c|}{$f (Hz)$} \\ \hline 
     & $C1$ & $C2$ & $C1$ & $C2$\\
    \hline
0	&	-	&	-	&	60.000	&	60.000	\\
1	&	-	&	-	&	59.991	&	59.991	\\
2	&	-	&	-	&	59.983	&	59.983	\\
3	&	-	&	-	&	59.976	&	59.974	\\
4	&	-	&	-	&	59.968	&	59.966	\\
5	&	-	&	-	&	59.960	&	59.957	\\
6	&	0.480	&	0.490	&	59.952	&	59.951	\\
7	&	0.470	&	0.500	&	59.944	&	59.941	\\
8	&	0.460	&	0.500	&	59.937	&	59.933	\\
9	&	0.460	&	0.500	&	59.930	&	59.924	\\
10	&	0.460	&	0.500	&	59.922	&	59.916	\\
11	&	0.450	&	0.490	&	59.915	&	59.908	\\
12	&\textbf{0.510}	&\textbf{0.530}	&	59.901	&	59.898	\\
 \hline
    \end{tabular}
\end{table}

\item {Case 2 ($C2$):} 
As can be seen from Table~\ref{tab:Evaluation parameters}, parameters $R,\ T$, and $AD$ remains the same as $C1$ while $H$ increases. In order to launch a successful attack, the attacker needs to be able to inject at least 6.0\% of false data into the measurements \ie $ToI$ = 6.0\% which results in $\Delta P_1^a$ = 1.015 p.u. This attack makes $\dot f$ reach 0.513 Hz/s in 12 cycles, which is greater than the $\overline{\dot f}$ of the generator at bus 4. Hence, the $ROCOF$ relay at bus 4 operates falsely and its generator gets disconnected from the grid ($P^{tg}$ = 1.0 p.u.). The behavior of frequency and $ROCOF$ are shown in Table~\ref{Tab:Output}.

\item Case 3 ($C3$): In this case, we show the number of successful FRO attacks from the 10,000 generated combinations and discuss the impact of the power system parameters on FRO attack success.
\end{itemize}

\begin{figure}[b]
    \centering
    \includegraphics[scale=0.4, keepaspectratio=true]{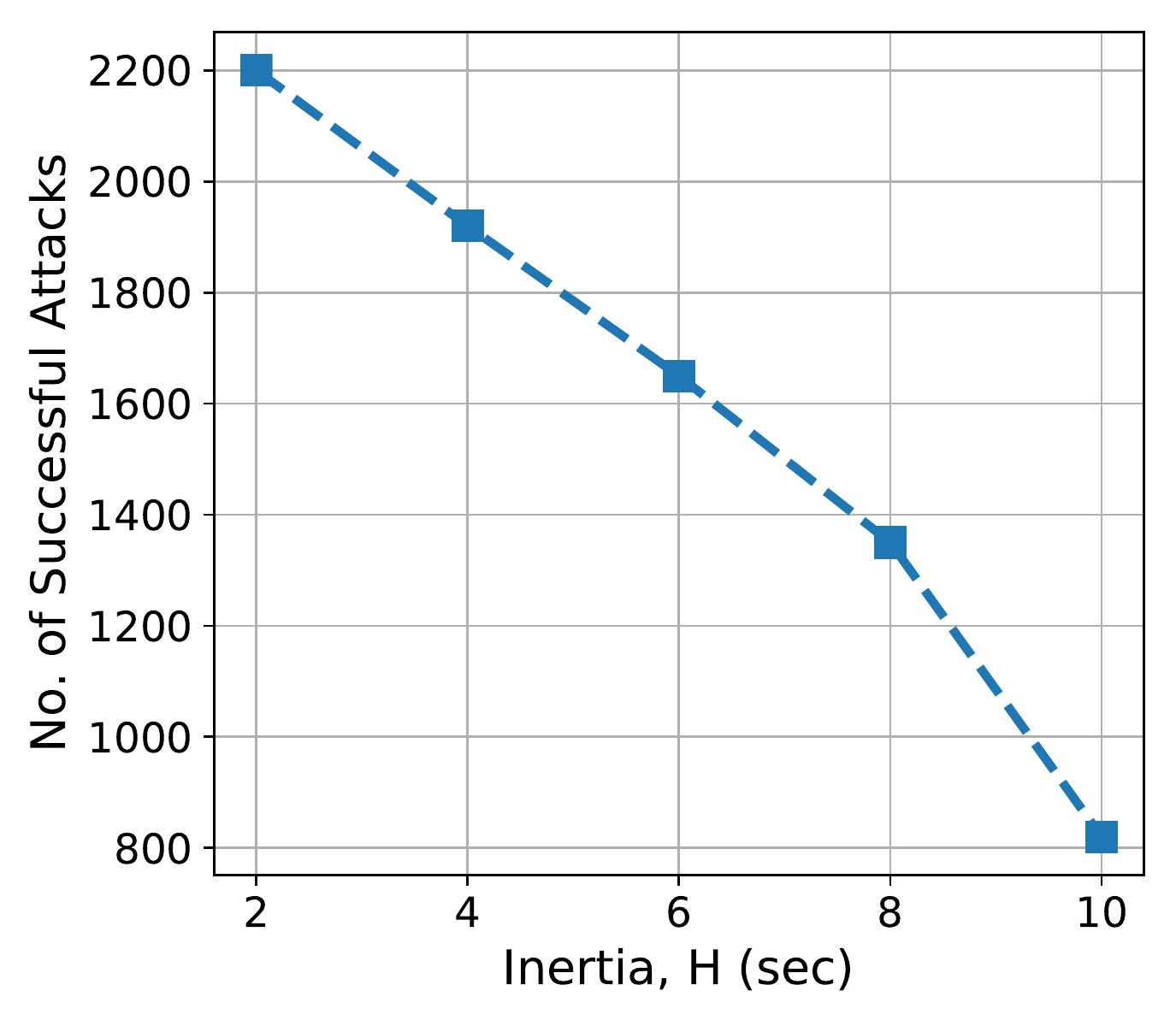}
    \caption{Number of successful FRO attacks vs. inertia ($H$).}
    \label{h_vs_ap}
\end{figure}

Fig.~\ref{h_vs_ap} shows the impact of $H$ on FRO success. As can be seen, there is almost a negative correlation between $H$ and the number of successful FRO. In other words, When $H$ increases, the number of successful FRO decreases. This is due to the fact that with a higher value of $H$, the frequency of the system becomes more stable and has less fluctuation. Therefore, the possibility of a larger $\dot f$ occurrence below $\overline{f}$ becomes less. 

\begin{figure}
    \centering
    \includegraphics[scale=0.4, keepaspectratio=true]{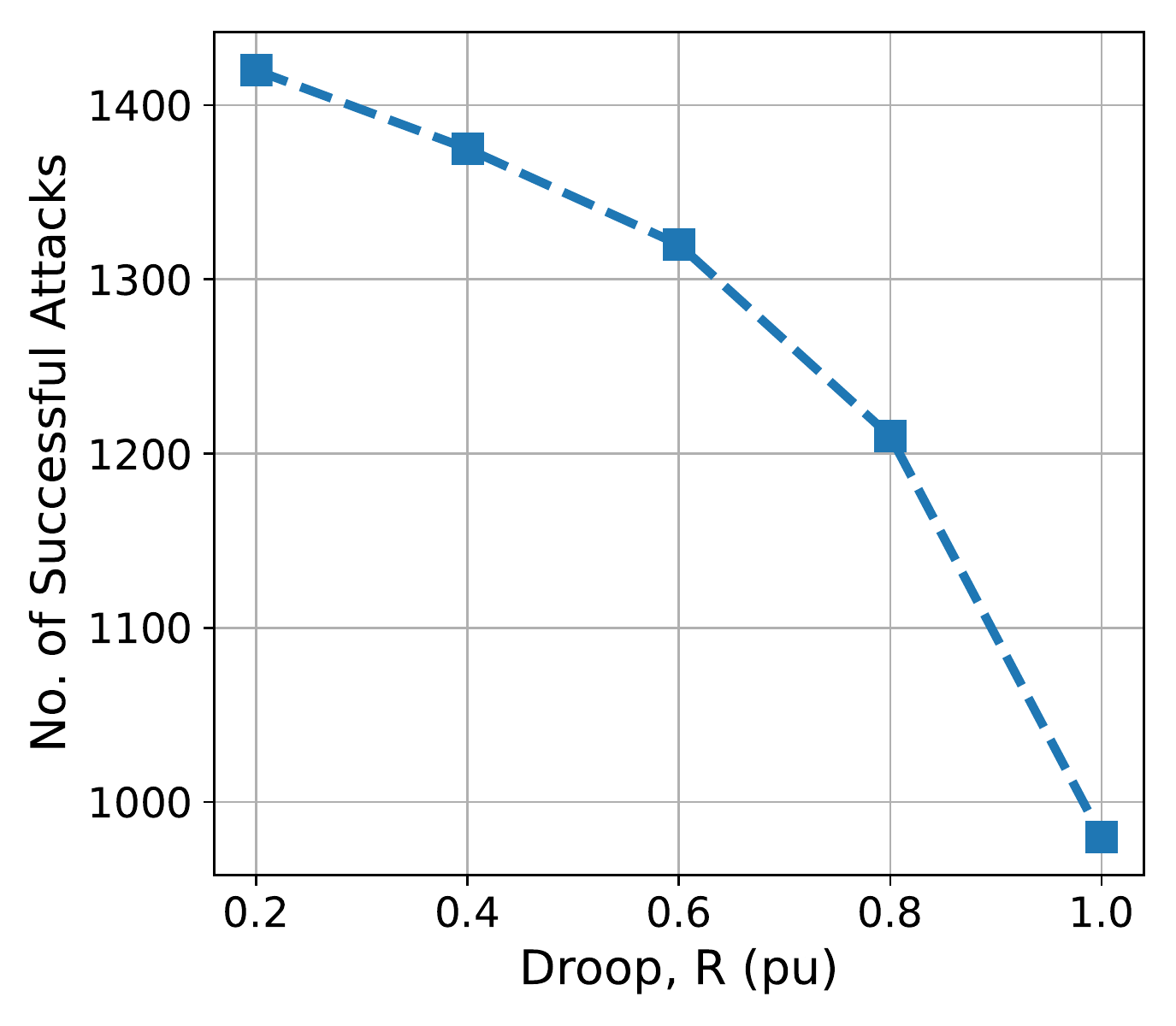}
    \caption{Number of successful FRO attacks vs.   governor's droop  ($R$).}
    \label{r_vs_ap}
\end{figure}

\begin{figure}
    \centering
    \includegraphics[scale=0.4, keepaspectratio=true]{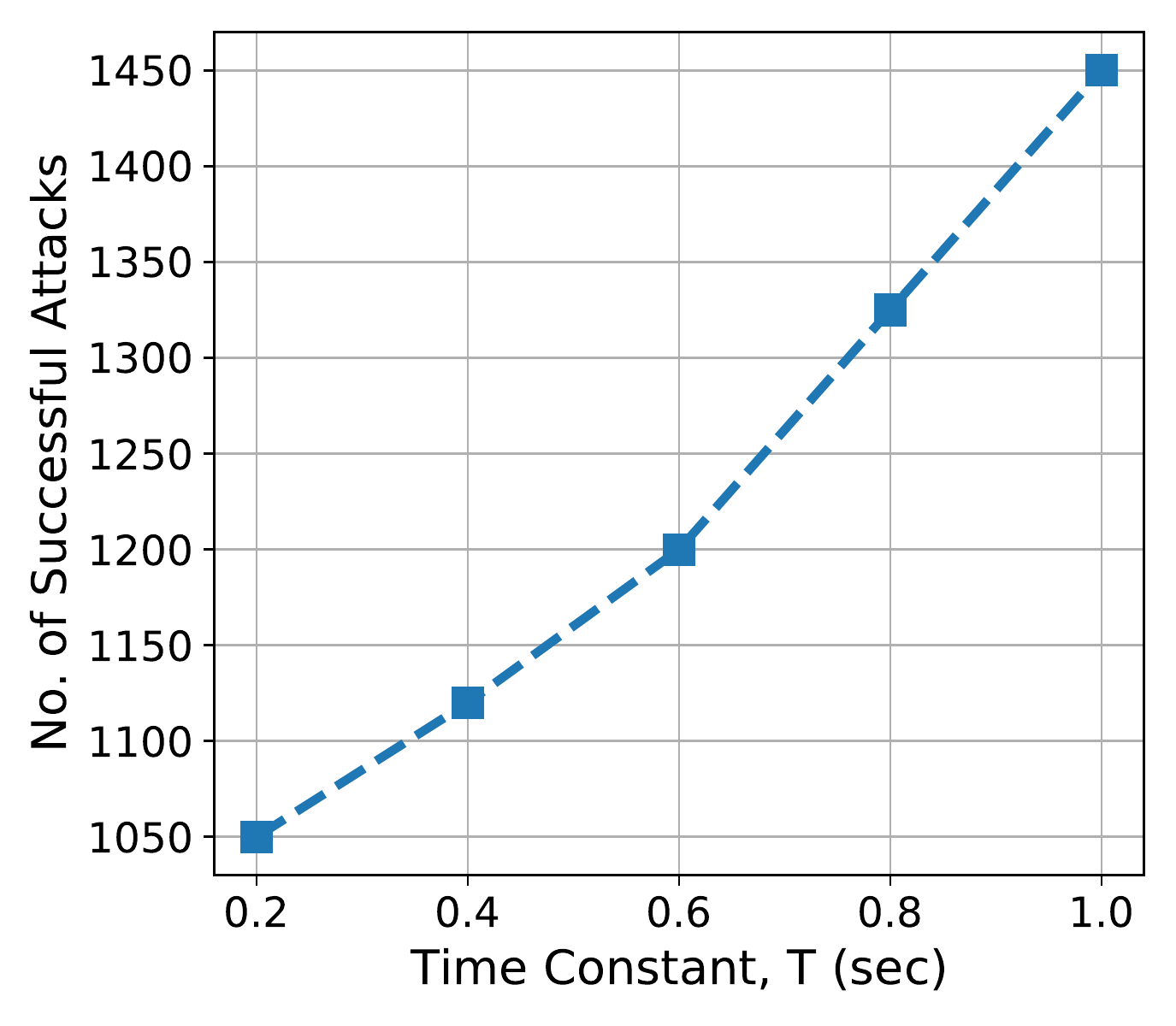}
    \caption{Number of successful FRO attacks vs. governor's time constant ($T$).}
    \label{t_vs_ap}
\end{figure}

Fig.~\ref{r_vs_ap} also shows that there is a negative correlation between $R$ and the number of successful FRO attacks. This is because, with increasing values of the droop, the power grid's generators show a slower reaction to the frequency changes in the power grid. This slow reaction goes against the attacker's goal, which tries to cause a sharp $\dot f$ or a frequency drop below $\overline{f}$. From Fig.~\ref{t_vs_ap}, it can be observed that an increase in $T$ leads to more successful FRO attacks. This positive correlation is due to the fact that larger $T$ makes the governor's response slower to any frequency abnormalities that actually gives more chances to the frequency fluctuations. This creates more possibilities for the attacker to achieve his goal. 

\begin{figure}
    \centering
    \includegraphics[scale=0.4, keepaspectratio=true]{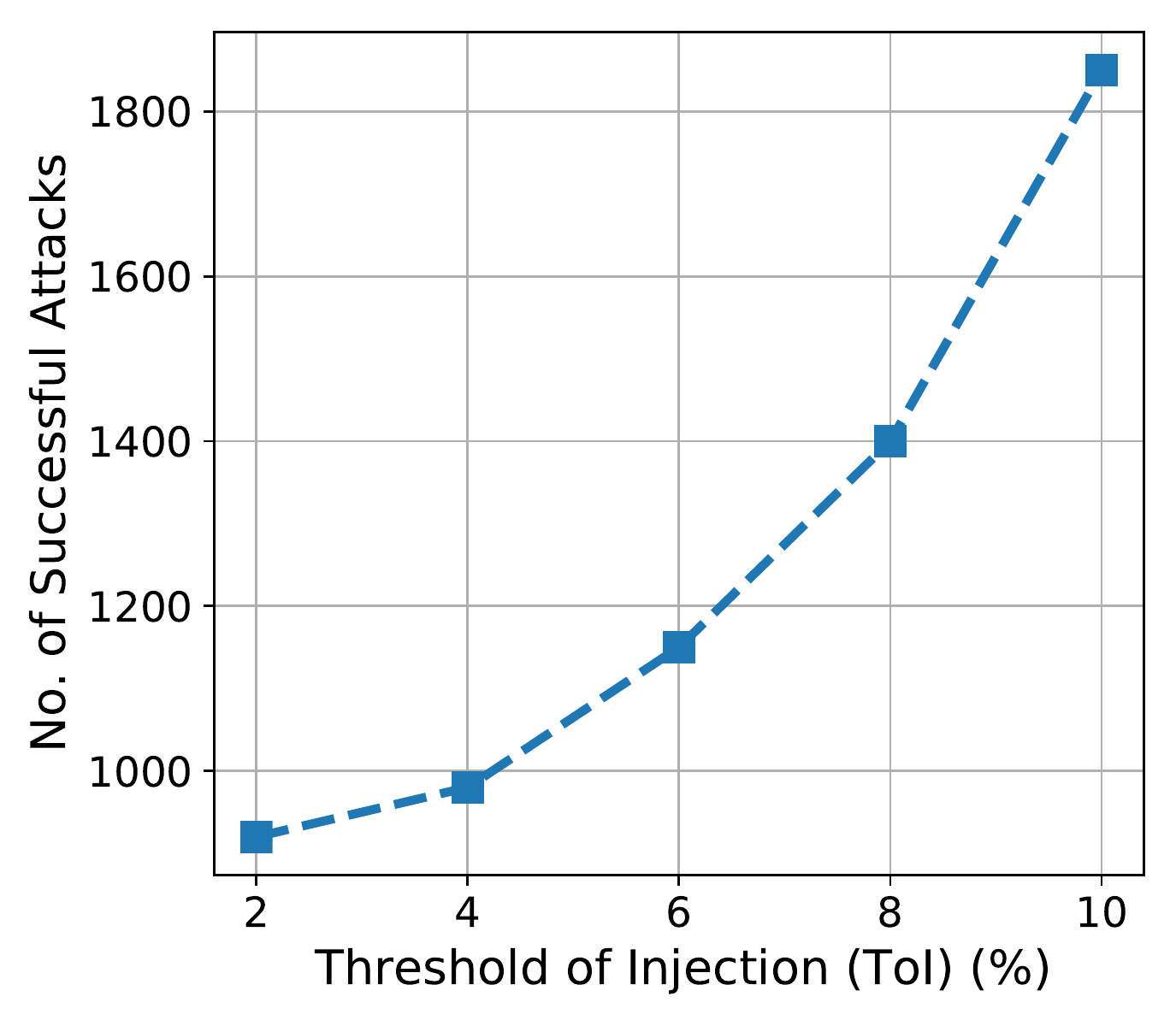}
    \caption{Number of successful FRO attacks vs. threshold of injection ($ToI$).}
    \label{sa_vs_ap}
\end{figure}

\begin{figure}
    \centering
    \includegraphics[scale=0.4, keepaspectratio=true]{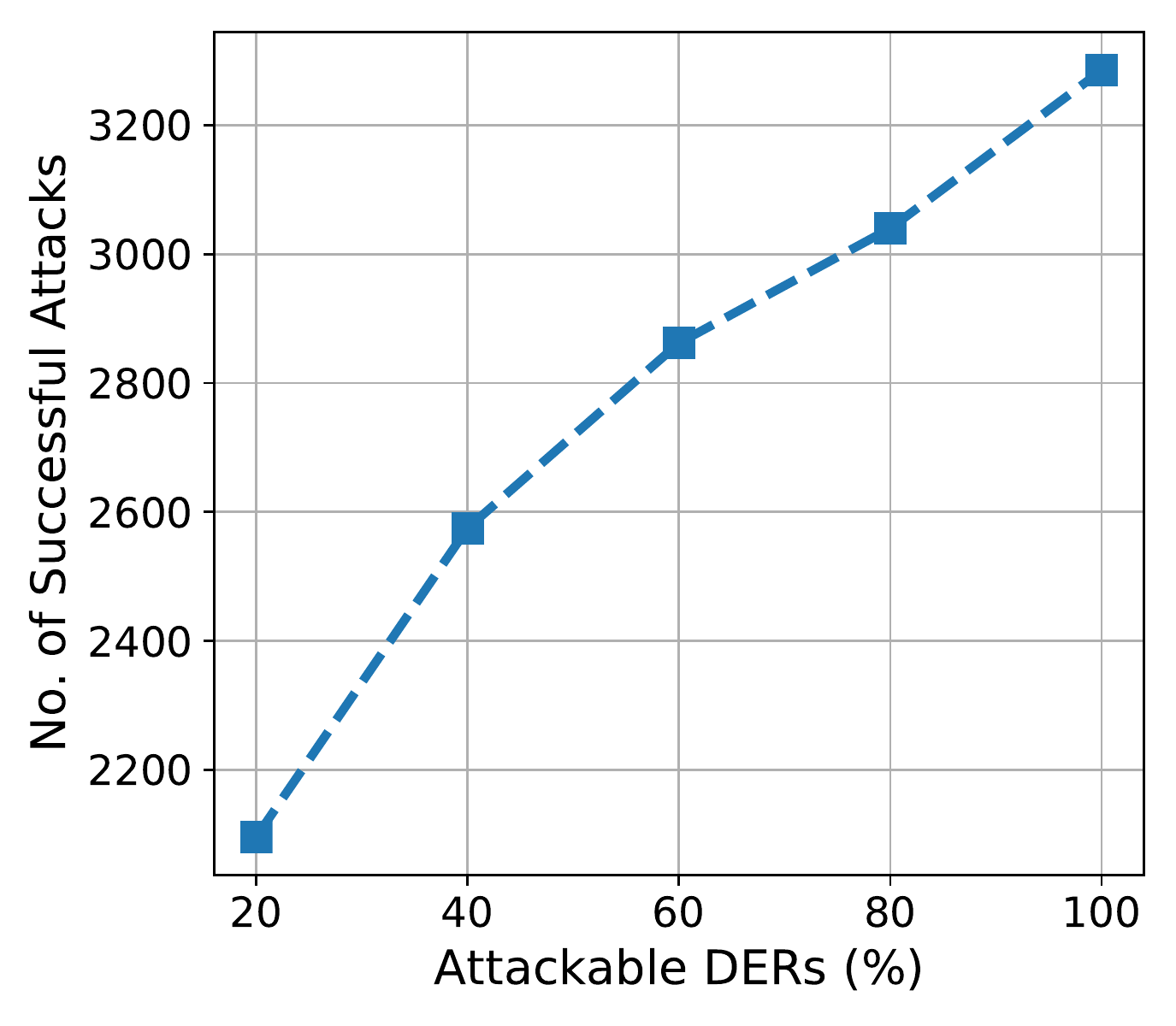}
    \caption{The number of successful FRO attacks vs. attackable DERs ($AD$). }
    \label{sp_vs_ap}
\end{figure}

In Fig.~\ref{sa_vs_ap}, it is shown that the relationship between $ToI$ and the number of successful FRO attacks is almost a proportional relationship. This is in accordance with the fact that if the attacker is able to change each of the measurements with a greater absolute value, the possibility of convincing the control center of sending a greater change to the generator setpoints is higher. The greater the changes in the setpoints, the greater the fluctuations in the frequency, and the greater the possibility of having successful FRO attacks. Fig.~\ref{sp_vs_ap}  the result of Fig.~\ref{sa_vs_ap}. It shows that if the attacker has access to a larger number of DER measurements, \ie greater $AD$, more successful FRO attacks are possible to happen in the grid.

Comparing Fig.~\ref{h_vs_ap}, Fig.~\ref{r_vs_ap},  and Fig.~\ref{t_vs_ap}, it can be seen that the slop of number of successful FRO attacks vs. $H$ is greater than the ones of $R$ and $T$. This shows the importance of considering the inertia of the power system while increasing the penetration of DERs in the grid.
Inertia can have significant effects on a power grid's vulnerability against FDI attacks. With increasing the penetration of low-inertia DERs such as wind turbines and solar panels into power systems, the frequency of the grid becomes less stable to any type of disturbances or attacks. This makes the power grids more potent to FRO attacks.

From Table~\ref{tab:Inertia impact on the attack type}, it can be observed that inertia has an impact on the FRO attack type (i. e., $ROCOF$ or $LS$ attack), as well. When $H$ is less than 4 (in the example power system), we have both $ROCOF$ and $LS$ attacks in the power grid while with an increase in $H$, we observe only $ROCOF$ attacks. 
This is because of the fact that for low $H$, the power grid shows a faster reaction to any load-generation imbalances. If these imbalances are big enough the frequency might drop in less than  6 cycles (the number of cycles considered in this paper for calculation of $\dot f$). This possibility of fast operation of $LS$ relays is eliminated with an increment of $H$.
Moreover, according to Fig.~\ref{r_vs_ap} and Fig.~\ref{t_vs_ap}, it can be noticed that 
an increase in $T$ or decrease in $R$ reduces the number of successful FRO attacks.

\begin{table}[h!]
    \centering
    \caption{Impact of inertia (H) on the attack type.}
    \label{tab:Inertia impact on the attack type}
    \begin{tabular}{|c|c|c|}
    \hline
        H (s) & $ROCOF$ Attack & $LS$ Attack \\ \hline
        $<$ 4 & \checkmark  & \checkmark  \\ \hline
        $\geq$ 4 & \checkmark   & $\times$\\ \hline
    \end{tabular}
\end{table}


\section{Conclusion}
\label{Sec:Conclusion}

In this work, we study the feasibility of FDI attacks on power systems that falsely trigger protective relays (i.e., rate-of-change-of-frequency and load shedding relays). 
The proposed formal model considers the impact of different power system's parameters, including generators' inertia, equivalent governors' droop and time constant, the threshold of false data injection, and the number of attackable DERs and synthesizes successful FRO attacks, if exists. Our results show that, among various parameters, inertia has the most impact on reducing the success of the FRO attacks as higher inertia can reduce the possibility of launching a successful FRO attack in the power system. Moreover, it is demonstrated that an increase in droop or decrease in the time constant of governors can lower the possibility of launching a FRO attack in the power system.


\bibliographystyle{IEEEtran}
\bibliography{sample-base}

\end{document}